\documentclass[%
prl,
twocolumn,
amsmath,amssymb,
aps,
]{revtex4-2}
\usepackage{amsmath}
\usepackage[normalem]{ulem}
\usepackage{tikz}
\usetikzlibrary{quantikz}
\usepackage[caption=false]{subfig}
\usepackage{graphicx} 
\usepackage{natbib}
\usepackage{bm}
\usepackage{amsthm}
\usepackage[normalem]{ulem}
\usepackage[unicode=true,pdfusetitle,
bookmarks=true,bookmarksnumbered=false,bookmarksopen=false,
 breaklinks=true,pdfborder={0 0 0},pdfborderstyle={},backref=false,colorlinks=true]
 {hyperref}

\newcommand{\bam}[1]{{\color{blue} B: #1}}

\renewcommand{\bam}[1]{}
\renewcommand{\section}[1]{{\it #1.}}
\renewcommand{\subsection}[1]{{\it #1. }}

\begin{document}

    \title{Pre-optimization of quantum circuits, barren
plateaus and classical simulability: tensor networks to unlock the variational quantum eigensolver}
	
\author{Baptiste Anselme Martin$^{1}$}%
\author{Thomas Ayral$^{2,1}$}
 \affiliation{1. 
 Eviden Quantum Lab, 78340 Les Clayes-sous-Bois, France}
 \affiliation{2. 
 CPHT, CNRS, Ecole Polytechnique, IP Paris, F-91128 Palaiseau, France
 }
	\begin{abstract}

        Variational quantum algorithms are practical approaches to prepare ground states, but their potential for quantum advantage remains unclear. Here, we use differentiable 2D tensor networks (TN) to optimize parameterized quantum circuits that prepare the ground state of the transverse field Ising model (TFIM). Our method enables the preparation of states with high energy accuracy, even for large systems beyond 1D. We show that TN pre-optimization can mitigate the barren plateau issue by  giving access to enhanced gradient zones that do not shrink exponentially with system size. We evaluate the classical simulation cost evaluating energies at these warm-starts, and identify regimes where quantum hardware offers better scaling than TN simulations.
	\end{abstract}
	
	\maketitle

	\begin{figure}
		\centering
        \includegraphics[trim=0.cm 0cm 4.cm 0cm, width=1.05\linewidth]{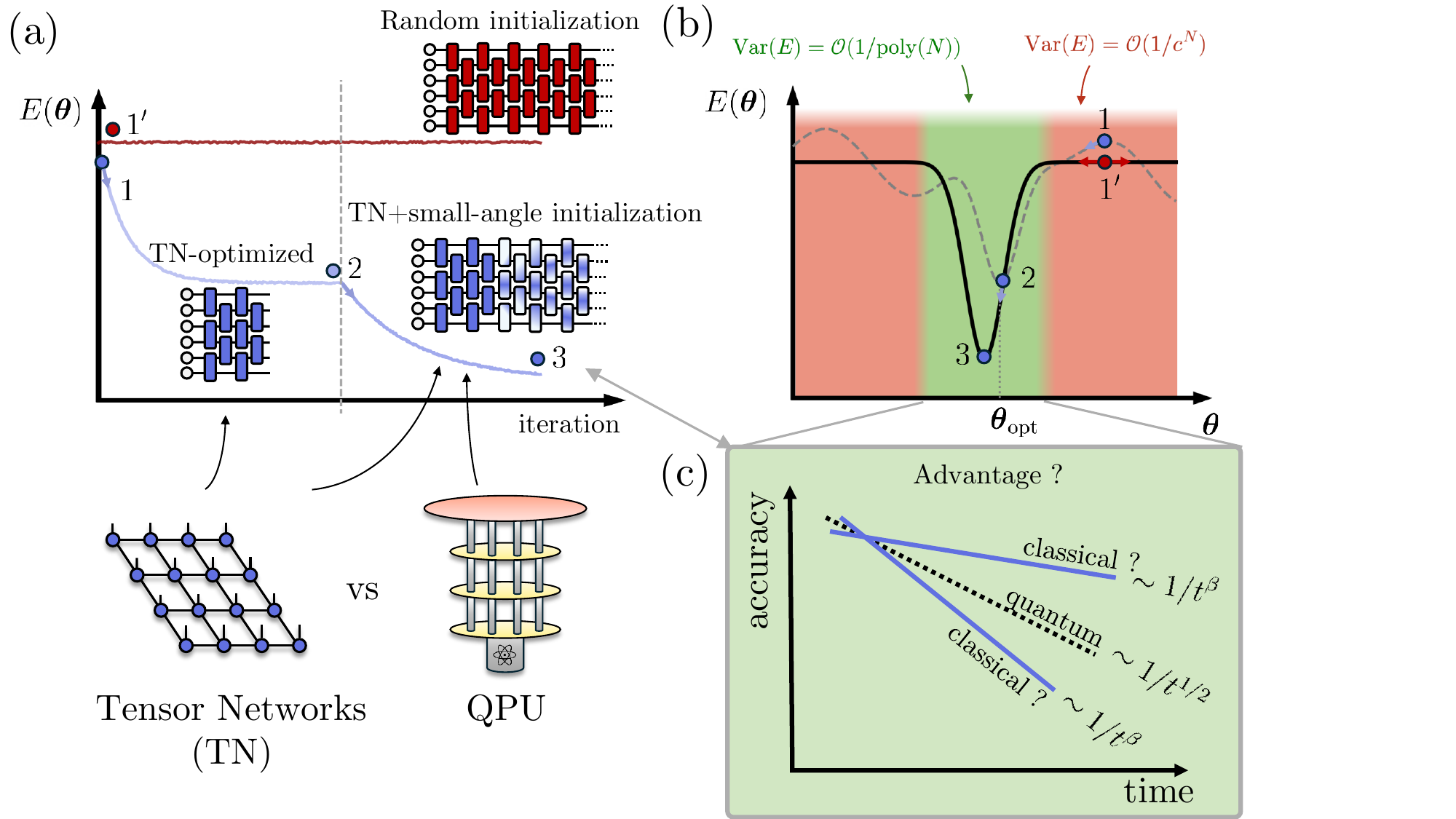}
		\caption{(a) A deep parameterized quantum circuit cannot be randomly initialized because of barren plateaus [1'], while a short-depth circuit can be optimized classically [1] and used as a warm-start for a deeper circuit [2], by initializing the remaining parameters around identity. (b) Our scheme allows to run VQE in regions with substantial energy  gradients and improve accuracy thanks to deeper circuits [3]. (c) We investigate the cost of evaluating the energy in the green region using TNs, and compare it to the sampling cost of a QC.}
		\label{fig:schemaglobal}
	\end{figure}

	Quantum hardware has been rapidly improving over the past years, bringing  hope that we could use quantum computing to simulate interacting many-body systems in regimes that have proven very challenging for classical algorithms. 
	Conversely, classical algorithms have also shown consistent progress, and analyzing their strengths and limitations provides valuable insights into the problems where quantum computers (QCs) could be most useful \cite{Ayral2025, waintal2026competequantumcomputerslecture}. A key example is the ground state (GS) search for weakly-entangled states, where classical techniques like tensor network (TN) methods, equipped with the density matrix renormalization group  (DMRG) \cite{PhysRevLett.69.2863} or imaginary-time evolution \cite{Paeckel_2019},
    can efficiently determine GSs.
    On the other hand, finding these GSs and their energies on QCs is challenging: fault-tolerant methods like quantum phase estimation \cite{1995quantum} \bam{Krylov too} require high-depth quantum circuits and initial states with significant overlaps with exact GSs \cite{Louvet2023, PRXQuantum.3.010318, ge2018fastergroundstatepreparation}, while near-term approaches like the variational quantum eigensolver (VQE) \cite{peruzzo2014variational, Tilly_2022} crucially depend on designing expressive {\it and} trainable  circuits.
    
    More precisely, a major challenge for variational quantum algorithms lies in avoiding barren plateaus (BPs) \cite{McClean_2018, Cerezo_2021, Harrow_2009, Larocca_2025}.  BPs refer to an exponentially fast flattening of the energy landscape with the system size, implying  an exponential number of measurements to carry out the optimization. However, BPs are only average properties of quantum energy landscapes. Recent works \cite{Holmes_2022,zhang2025escapingbarrenplateaugaussian,Puig_2025,mhiri2025unifyingaccountwarmstart} indicate the existence of regions in the parameter space (dubbed \textit{fertile valleys}) with non-vanishing gradients, which allow for an optimization, and whose size decays at worst only polynomially with system size.
    Initializing parameters in these regions would enable the optimization of quantum circuits while avoiding BPs.
    Yet, these regions often arise from restricting the computational problem to effective subspaces that scale polynomially with system size, which in turn makes the optimization classically simulable at polynomial cost \cite{Cerezo_2025, lerch2024efficientquantumenhancedclassicalsimulation}.

In this work, we explore the combination of classical TN algorithms with quantum computing for GS preparation and parameter initialization in variational quantum algorithms.
More specifically, we leverage two-dimensional TNs, called projected entangled pair states (PEPS \cite{Or_s_2014}), along with automatic differentiation \cite{Liao_2019}, to optimize parameterized quantum circuits (PQCs) for finding GSs of model Hamiltonians.
The resulting optimized quantum circuits can be used as initial states for further quantum computations, but also as \textit{warm-start} initial parameters for VQE using deeper quantum circuits, aiming for better accuracy while avoiding BPs.
After presenting the technical details and the performance of PEPS to optimize PQC, we demonstrate that our scheme allows us to initialize deeper quantum circuits in trainable regions with substantial gradients whose sizes and magnitudes decay at worst polynomially with system size and circuit depth. Finally, we address the classical simulability of this protocol:  we quantify the cost of TN simulations and compare it to the sampling cost on a QC, and identify cases where QCs could bring a polynomial advantage over classical TN simulations within the framework of variational quantum algorithms (see Fig. \ref{fig:schemaglobal}).

    \bam{Algo NISQ toujours utile meme en FT a priori ? Ici on regarde que QC noiseless.}

    \begin{figure}
        \centering
        \includegraphics[width=1.\linewidth]{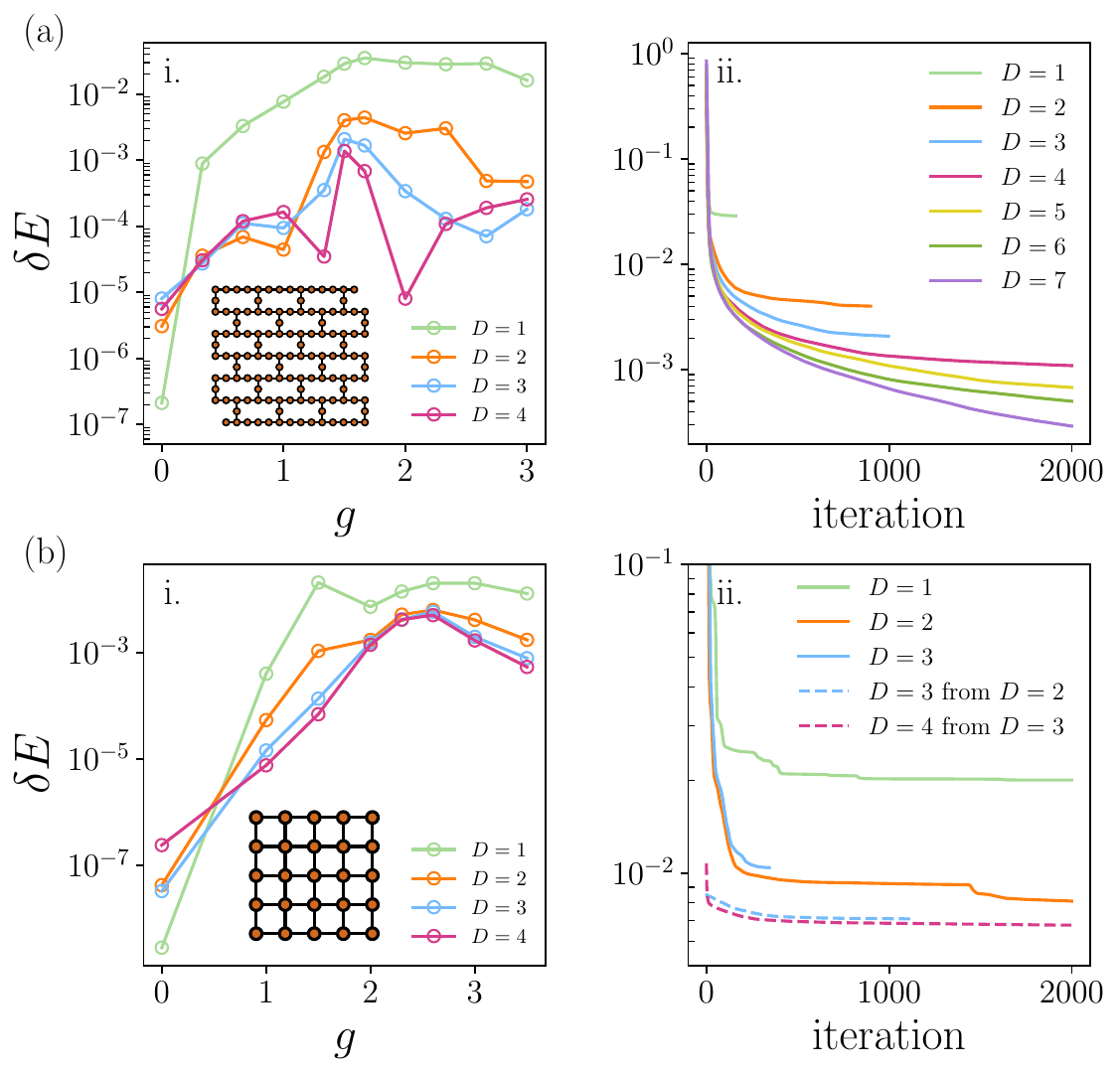}
        \caption{\emph{Energy convergence of PEPS optimization for the TFIM.} (a) 127 qubit heavyhex lattice: i. Relative energy error $\delta E$ vs $g$ for different circuit depths $D$.  ii. $\delta E$ vs. the number of iterations for different circuit depths $D$ at $g_c \simeq 1.5$, and $\chi=8$.  (b) 5$\times$5 square lattice: i. $\delta E$ vs $g$ for different circuit depths $D$. Here the energy is computed by statevector simulations obtained from the TN optimizations. ii. $\delta E$ vs the number of iterations and circuit depth $D$ at $g_c \simeq 2.6$, and $\chi$ up to 6 with SU-like expectation values.}
        \label{fig:figure_vqe}
    \end{figure}

    \section{Training VQE with PEPS}\label{sec:vqe_peps} Integrating classical algorithms with quantum computing appears to be a promising practical  approach. In particular, TNs have proven  competitive to simulate QCs \cite{Zhou_2020, Ayral_2023, Noh_2020, Zhang_2022, Tindall_2024, Begu_i__2024, Seitz_2023, Patra_2024,müller2024enablinglargedepthsimulationnoisy,pang2020efficient, Lee_2025}
    and even enhance quantum computations, whether to encode TN states \cite{PhysRevLett.95.110503, Ran_2020, PRXQuantum.5.030344,PhysRevA.101.010301,rudolph2022decompositionmatrixproductstates,jamet2025anderson, jaderberg2025variationalpreparationnormalmatrix}, compress quantum circuits \cite{lin_real-_2021,PhysRevA.109.062437,Tepaske_2023, Keever:2022hfy, keever2023adiabatic,causer2023scalable, le2025riemannianquantumcircuitoptimization, gibbs2025deep}, or optimize PQCs for GS preparation \cite{dborin2021matrixproductstatepretraining,PhysRevX.12.011047, Rudolph_2023, PhysRevResearch.5.033141, khan2023preoptimizingvariationalquantumeigensolvers, rogerson2024quantumcircuitoptimizationusing, PRXQuantum.6.010320}.
    A key constraint in TN simulations is the choice of network topology. While one-dimensional TNs, called matrix product states (MPS) \cite{Schollw_ck_2011,Or_s_2014, Cirac_2021}, and tree tensor networks (TTN) \cite{Shi_2006, PhysRevLett.126.170603, Seitz_2023} offer well-controlled approximations and manageable contraction costs at the price of handling well only low-dimensional problems, networks with loops, such as PEPS \cite{Or_s_2014, Cirac_2021}, can handle larger dimensions, but cannot be contracted efficiently unless approximately---and still at a high polynomial cost \cite{verstraete2004renormalizationalgorithmsquantummanybody,Lubasch_2014}. 
	
	Here, we employ PEPSs to capture states with intrinsic correlations that are beyond 1D. More specifically, we choose PEPS TNs combined with the so-called simple update (SU) algorithm \cite{Shi_2006,Jiang_2008, Lubasch_2014,Jahromi_2019,Tindall_2024, Patra_2024} to represent the state obtained by a PQC $U(\bm{\theta})$. Gate application is done at cost $\mathcal{O}(\chi^{d+1})$, with $d$ the network coordination number. This leads to a PEPS approximation $|\Psi_{\text{PEPS}  }(\bm{\theta}) \rangle \simeq U(\bm{\theta}) |\Psi_0 \rangle$. The energy expectation value $E(\bm{\theta})$ can then be computed either in SU-fashion at cost $\mathcal{O}(\chi^{d})$ \cite{Tindall_2024,Jahromi_2019, Patra_2024}, or via a MPS-boundary contraction \cite{verstraete2004renormalizationalgorithmsquantummanybody,Lubasch_2014}. The former is particularly advantageous for networks with large loops and tree-like correlations, allowing for large bond dimensions, while the latter is better suited for more connected networks like square grid lattices.
	The gradients required by SciPy's L-BFGS-B optimizer \cite{2020SciPy-NMeth} are obtained by automatic differentiation, avoiding the costly, parameter-wise gradient evaluations required by parameter shift rules
     or any other finite difference evaluation \cite{Schuld_2019} on QCs.

	To benchmark the capabilities of PEPS+SU for optimizing GS quantum circuits, we focus on the transverse field Ising model (TFIM), whose Hamiltonian
    \begin{equation}
		H = -\sum_{\langle i,j \rangle} Z_i Z_j - g\sum_i X_i
	\end{equation}
    describes nearest-neighbor interactions between spins along the $z$ direction competing with a transverse magnetic field $g$.
   	This paradigmatic model exhibits a phase transition at a lattice-dependent critical value $g_c$. 
    As the capabilities of PEPS+SU algorithms strongly depend on the topology of the graph and the importance of loop correlation versus local tree-like correlations, we focus on different graphs: the heavyhex topology used in superconducting qubit prototypes \cite{kim2023evidence} and the 2D square lattice.

	We choose brickwall circuits composed of SO(4) gates, parameterized by 6 parameters, as in \cite{PhysRevX.12.011047}, applied only on neighboring qubits, and with a variable depth $D$ (see Fig. \ref{fig:schemaglobal} (a)). To mitigate truncation errors, we use a \textit{SU-regauging} \cite{Tindall_2023,Patra_2024, PhysRevResearch.3.023073} applied after each brickwall layer of the circuit by effectively running the same brickwall layer backwards with all gates replaced by identity gates. For both lattices, we compute expectation values in a SU-type fashion.
    
	Fig. \ref{fig:figure_vqe} (a) shows the results obtained for the 127-qubit heavyhex topology with a bond dimension $\chi = 8$, with reference energies obtained from a converged imaginary time-evolution of PEPS+SU. In Fig \ref{fig:figure_vqe} (a) i. we perform the optimization procedure for a range of values for $g$ and different depths, allowing to prepare GSs with good energy accuracy.
    We then focus on the critical point $g = g_c \simeq 1.5$ in Fig \ref{fig:figure_vqe} (a) ii.  Here, we optimize deeper circuits and find that increasing the circuit depth consistently helps find lower energy states, despite keeping the bond dimension $\chi$ low, showing the strength of using a tailored network topology.

In contrast, the square lattice appears to be much more challenging.
 In Fig. \ref{fig:figure_vqe} (b) we show that we are still able to optimize short-depth quantum circuits within these approximations, especially away from $g_c$.
For deeper circuits, however, we observe that initialization plays a key role: with a depth $D = 3$ quantum circuit, TN approximations (SU update and SU-type expectation value computations) cause the optimization from randomly initialized set parameters to fail. 
This can be solved by choosing initial parameters from previously optimized $D=2$: then, we can optimize the quantum circuits to lower-energy states (same from $D=3$ to 4).

    \begin{figure*}
		\centering
		\includegraphics[trim=0.cm 0cm 0.cm 0cm,width=1.\linewidth]{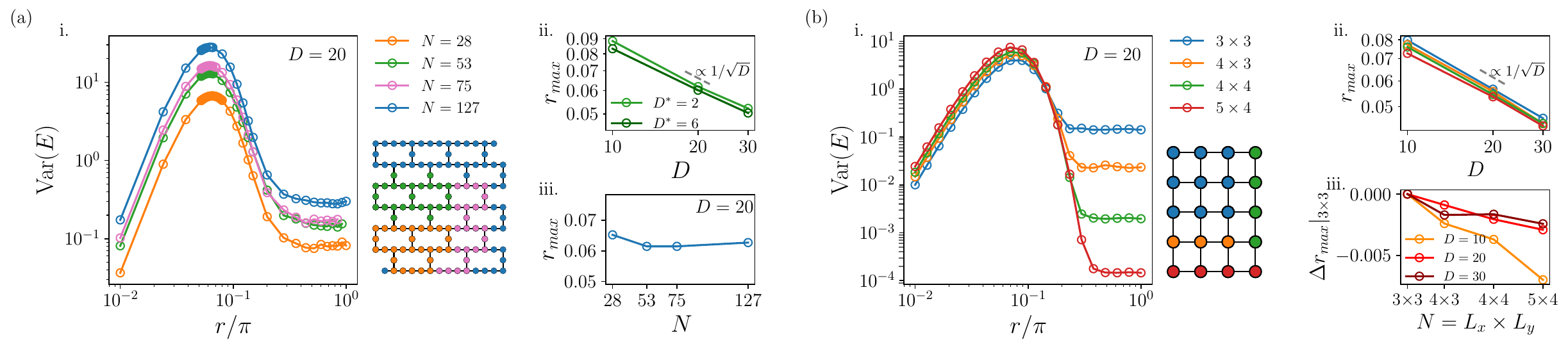}
		\caption{\emph{Energy landscape diagnostic}. (a) Heavyhex lattice including (i) $\text{Var}(E)$ vs $r$ for $N = 28, 53, 75$ and 127, $r_{\text{max}}$ vs (ii) total depth $D$ with $D^*=2$ and $6$  and (iii) vs system size for $D^*=2$ and $D=20$ on the 53-qubit system. (b) Square lattice with (i) $\text{Var}(E)$ vs $r$ for $N=$ $3\times 3$, $4\times 3$, $4\times 4$ and $5\times 4$,  $r_{\text{max}}$ vs (ii) total depth $D$ with $D^*=2$ and (iii) $r_{\text{max}}$ vs the system size, relative to $r_{\text{max}}$ for $N= 3\times 3$. }
		\label{fig:bp_vs_size} 
	\end{figure*}

\section{Avoiding barren plateaus with PEPS warm-starts}\label{sec:bp}        
We demonstrated that PEPS can effectively optimize quantum circuits for GS preparation. However, optimizing deep circuits remains difficult for TNs due to entanglement growth.  Continuing the optimization on a quantum device may thus be necessary. As noted earlier, parameter initialization is crucial for VQE performance: poor initialization can produce quantum states with exponentially small gradients, preventing efficient optimization, as illustrated in Fig. \ref{fig:schemaglobal}.

	    A natural strategy is to use classical simulations to pre‑optimize quantum circuits, providing a \textit{warm-start} initialization that helps avoid BPs \cite{Rudolph_2023,Goh_2025, PhysRevResearch.4.033012,rad2022survivingbarrenplateauvariational,ravi2023cafqaclassicalsimulationbootstrap,cheng2022cliffordcircuitinitialisationvariational,dborin2021matrixproductstatepretraining, khan2023preoptimizingvariationalquantumeigensolvers}. In this work, we examine how shallow circuits optimized with tensor networks can be embedded within deeper quantum circuits. We use techniques developed in \cite{Holmes_2022,Puig_2025,mhiri2025unifyingaccountwarmstart} to probe the energy landscape around the warm-start point. More specifically, we consider a quantum circuit of depth $D$, where the first $D^*$ layers are optimized with TNs, the rest of the parameters being set to 0. These initial parameters are denoted $\bm{\theta}_\text{opt} $. We then draw 1000 samples uniformly in the hypercube $[ \bm{\theta}_\text{opt} - r, \bm{\theta}_\text{opt} +r]$, where $r$ is the size of the hypercube in the parameter space around  $\bm{\theta}_\text{opt}$, and compute the energy variance $\text{Var}(E)$. By varying $r$, we make an energy landscape diagnostic around $\bm{\theta}_\text{opt}$: the energy variance acts here as a trainability metric as it reflects the gradient magnitude. 
    
    Fig. \ref{fig:bp_vs_size} shows the energy variance as a function of the distance $r$ in parameter space from the initial point for the (a) heavyhex and (b) square lattices.
	For the square lattice, we explore the dependency of $\text{Var}(E)$ on $r$ for $3\times3, 4\times3, 4\times4$ and $5\times4$ lattices as well as different depths $D$ via exact statevector simulations, with a warm-start point from $D^*=2$.
    At large $r$, we observe the BP phenomenon, with an energy variance decaying exponentially with system size. In contrast, around the warm-start point $r =0$, there exists a region with energy gradients orders of magnitude larger than those of randomly selected samples.
    As predicted in \cite{mhiri2025unifyingaccountwarmstart}, we observe that the size of the region with sizable gradients, whose proxy is the $r$ with the largest variance, dubbed $r_{\text{max}}$, does not decrease exponentially with system size and depth, but rather linearly with system size and proportionally to $1/\sqrt{D}$. 
	
    The more TN-friendly heavyhex topology allows us to explore larger system sizes than  the square lattice, as well as the impact of the number of optimized layers $D^*$. Fig. \ref{fig:bp_vs_size}(b) shows the energy variance as a function of $r$ for $N = 28,53,75$ and $127$. The optimized points are obtained using $\chi = 8$, and the sampling with $\chi = 16$. In this setting, $r_{\text{max}}$ is efficiently identified, with converged results in $\chi$. We find a trainable region around the PEPS optimized warm-starts, whose size  hardly depends on the system size. For larger $r$'s, we could not converge in bond dimension due to the larger number of qubits; we thus cannot observe the exponential decay characteristic of BPs.

	\section{Classical simulability}\label{sec:class_sim}
	So far, we have shown that pre-optimizing parameterized quantum circuits combined with a small-angle initialization provides a viable path to avoid BPs. 
    However, the question of the classical simulability of this strategy remains: while deep circuits are challenging to simulate for TNs,  the region of non-vanishing gradients shrinks  (at worst polynomially) with the number of qubits and circuit depth, which lowers the expressivity and entanglement power of the quantum circuits. 
    Since we are exploring a restricted subspace only, is it worth using a quantum processor? Is there room for a polynomial advantage?

    To answer these questions, we assess the time complexity of evaluating energy expectation values with TN simulations as a function of accuracy and compare it to the sampling cost of the equivalent quantum computation. Let us first turn to the energy error as a function of simulation time. Since the simulation time is implementation-dependent, we focus instead on its scaling. This classical scaling is compared with the scaling of the quantum computing time $t$, itself proportional to the number of shots $M$. The QC error comes from the sampling, and scales as $\epsilon_{\text{QC}} \propto 1/\sqrt{M} \propto 1/\sqrt{t}$ \cite{Wecker2015}, even using advanced techniques like classical shadows \cite{Huang2020}.
	
	To evaluate the error scaling of the TN simulation, we consider different optimized depths $D^*$ and draw 10 random parameters taken in the hypercube of size $r_{\text{max}}$ \bam{préciser combien}. We evaluate the energy $E(\boldsymbol\theta)$ with TN simulations of different costs (i.e. different bond dimensions), and compute its error $\epsilon_{\text{TN}}$ with respect to an exact solution or a converged TN simulation. 
    We then average the errors corresponding to the same bond dimensions and extract the scaling of the accuracy as a function of time by fitting $\epsilon_{\text{TN}} = \alpha /t^{\beta}$.
    $\alpha$ is an overhead factor and $\beta$ a scaling exponent that links accuracy with time. $\beta$ is intrinsic to the algorithms used here, and independent of the implementation and (classical) hardware.
    We then compare $\beta$ to the quantum scaling sketched above, $\beta_{\text{QC}} = 1/2$.
    For a given circuit topology, system size $N$, optimized depth $D^*$ and total depth $D$, we can then claim a potential quantum advantage if the scaling of the TN simulation is worse than the quantum sampling cost, i.e if $\beta <\beta_{\text{QC}} = 1/2$.

	\begin{figure*}
		\centering
		\includegraphics[width=1.\linewidth]{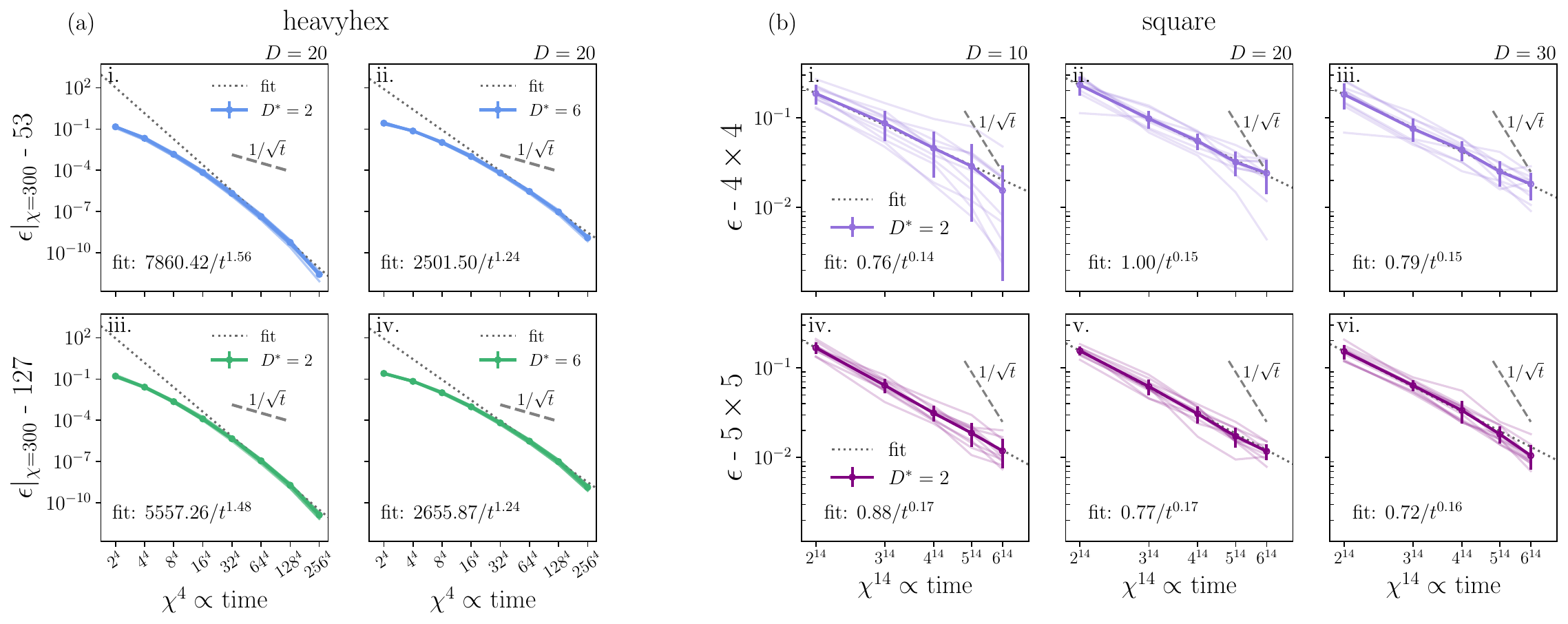}
		\caption{\emph{Average error $\epsilon$ versus time in the trainable region of  size $r_{\text{max}}$}. PQCs of depth $D$ pre-optimized with  a $D^*$ quantum circuit. (a) Heavyhex lattice with (i) 53 qubits with  $D^* = 2$ and (ii) $D^* = 6$ and (iii) 127 qubits with $D^* = 2$ and (iv) $D^* = 6$. (b) Square lattice with system sizes (a) $4\times 4$ and (b) $5 \times 5$ with $D^* = 2$ and (i, iv) $D = 10$,  (ii,v) 20 and (iii, vi) 30. Dashed lines: quantum sampling error.}
		\label{fig:dE_vs_time}
	\end{figure*}

Fig. \ref{fig:dE_vs_time}(a) shows results obtained for the heavyhex topology with $N = 53$ and $127$. For both system sizes, we evaluate $\epsilon_{\text{TN}}$ for a circuit of depth $D=20$ with a pre-optimized depth $D^* = 2$ (i, ii) and $6$ (iii, iv).
For all settings, the size $r_{\text{max}}$ of the enhanced gradient region is evaluated using the same method as before.
For this topology, despite considering a depth that is in general challenging to simulate, the TN simulations perform particularly well in the region of trainability defined by $r_{\text{max}}$, with a measured scaling $\beta \approx 1.56-1.48$ much more favorable compared to the quantum scaling $\beta_\mathrm{QC} = 1/2$.
Interestingly, this observation still holds when putting a larger load on the TN computation: 
when increasing $D^*$ from 2 to 6, we reduce the part of the quantum circuits initialized around identity, and thus in principle raise the bar for TNs.
While we do observe a slight increase in the error $\epsilon_{\text{TN}}$, the scaling factor $\beta$ only decreases from 1.54 to 1.24.
This slow decrease can be explained by the fact that a lower $D^*$ induces a decrease of $r_{\text{max}}$, and thus makes the rest of the quantum circuits less entangling. As for the size dependence, we observe no clear change in $\beta$ when going from 53 to 127 qubits. 

We finally explore, in Fig.\ref{fig:dE_vs_time}(b), the case of a square lattice system.
Its high connectivity and short-loop correlations render the SU-type expectation values less accurate, warranting more sophisticated contraction schemes. While the evolution of the TN state is performed within the SU approximation using bond dimensions $\chi = 2, 3, 4, 5,$ and $6$, we evaluate each term $P$ in the Hamiltonian $H$ using the MPS-boundary algorithm, which contracts $\langle \Psi | P | \Psi \rangle$ with the help of an environment MPS of bond dimensions $\chi_E = 4, 9, 16, 25,$ and $36$ (the contraction cost is $\mathcal{O}(\chi^8 \chi_E^3)$ \cite{ pang2020efficient}). 
Due to the high cost of TN contraction, we restrict our study to small system sizes: $4 \times 4$ (i–iii) and $5 \times 5$ (iv–vi), allowing us to compute exact expectation values via statevector simulations. On the other hand, this enables us to explore higher total depths $D$. The scaling factor is fitted using the points where $\chi_E = \chi^2$, which appears to be a good trade-off between computational time and accuracy. For the settings studied here, we observe little dependency on $D$ and system size, suggesting that the reduction of $r_{\text{max}}$ with these parameters counterbalances the increase in computational cost. 
\bam{This sentence could be clarified further—perhaps rephrase to emphasize the trade-off between parameter space size and simulation cost.}
However, we find a scaling factor $\beta \simeq 0.15 - 0.17< \beta_{\text{QC}} = 1/2$, pointing to a polynomial advantage of a QC over TN.

These examples highlight that, despite reaching parameter regions with non-vanishing energy differences---and thus trainable ones---the question of whether it is advantageous to continue a variational quantum optimization on a QC is subtle.
We show here that, as suggested in \cite{Cerezo_2025}, these regions are classically simulable, i.e., the simulation error $\epsilon_{\text{TN}}$ decreases polynomially with time.
However, not all topologies are equal from a TN perspective, and we found that a polynomial quantum advantage over TNs can be reached in variational quantum algorithms while avoiding barren plateaus only for some topologies.

We anticipate that moving towards more challenging problems---such as more connected lattices, longer-range interactions, and more entangled GSs---should lead to stronger advantages than those uncovered here.
To make a fair comparison, one should also factor in the cost of gradient evaluation: while computing gradients through TNs can benefit from backpropagation and is thus weakly dependent on the number of parameters (though the memory overhead caused by automatic differentiation might represent a bottleneck when using large bond dimensions \cite{Liao_2019}), computing gradients on QCs comes with a cost proportional to the number of parameters, possibly negating the above advantage.

\section{Discussion} BPs come from the exponential decay of the \emph{average} variance of the energy when using very expressive circuits. 
Yet, it does not rule out the existence of "fertile valleys", namely non-exponentially shrinking regions with non-vanishing gradients that can lead to a minimum. \bam{(or at least a local)} 
This paper addresses two questions: (i) can one find a classical method to initialize the variational search in a "fertile valley"? and (ii) once in this valley, is there something to gain from using a QC to continue the optimization?
We answer these questions in the paradigmatic case of the TFIM in 2D on two lattice geometries for sizes up to 127 qubits.
Our answer to question (i) is yes: PEPSs allow us to find regions of non-vanishing gradients whose size does not decrease exponentially with system size.
As for question (ii), we show that once in the valley, both quantum and TN methods come with polynomial costs but with different exponents, and that quantum methods supersede TNs for more connected lattices.

Last, we emphasize that the notion of quantum advantage is here only defined within the framework of variational GS preparation on a QC: we have tackled the question of whether we should use a TN or a QC to optimize a quantum circuit to prepare a GS.
From the sole perspective of GS search, other classical algorithms should be compared---as done e.g. in \cite{PhysRevX.12.011047, miao2025convergence}, which indicates a potential advantage of classes of parameterized quantum circuits over TN algorithms.
		
\begin{acknowledgments}

{\it Acknowledgments.} This work is part of HQI initiative (www.hqi.fr) and is supported by France 2030 under the French National Research Agency award number “ANR-22-PNCQ-0002.  TA is supported by France 2030 under the French National Research Agency award number ANR-22-EXES-0013.
\end{acknowledgments}

\bibliography{biblio}

\end{document}